\documentclass[sigconf]{acmart}

\AtBeginDocument{%
  \providecommand\BibTeX{{%
    \normalfont B\kern-0.5em{\scshape i\kern-0.25em b}\kern-0.8em\TeX}}}

\setcopyright{acmcopyright}
\copyrightyear{2022}
\acmYear{2022}
\acmDOI{10.1145/3491101.3519724}

\acmConference[CHI '22 Extended Abstracts]{Extended Abstracts of the CHI 2022 Conference on Human Factors in Computing Systems}{Apr 30-- May 05,
  2022}{New Orleans, LA}
\acmPrice{15.00}
\acmISBN{978-1-4503-XXXX-X/18/06}
\acmBooktitle{Extended Abstracts of the CHI 2022 Conference on Human Factors in Computing Systems}

\usepackage{multicol}
\usepackage{mdframed}

\begin{document}

\title{Prescriptive and Descriptive Approaches to Machine-Learning Transparency}

\author{David Adkins}
\email{davidadkins@fb.com}
\affiliation{%
  \institution{Meta AI}
  \country{USA}
}
\author{Bilal Alsallakh}
\authornote{Equal contribution (authors' list in alphabetical order).}
\email{bilalsal@fb.com}
\affiliation{%
  \institution{Meta AI}
  \country{USA}
}
\author{Adeel Cheema}
\email{acheema@fb.com}
\affiliation{%
  \institution{Meta AI}
  \country{USA}
}
\author{Narine Kokhlikyan}
\email{narine@fb.com}
\affiliation{%
  \institution{Meta AI}
  \country{USA}
}
\author{Emily McReynolds}
\email{emcr@fb.com}
\affiliation{%
  \institution{Meta AI}
  \country{USA}
}
\author{Pushkar Mishra$^{*}$}
\email{pushkarmishra@fb.com}
\affiliation{%
  \institution{Meta AI}
  \country{UK}
}
\author{Chavez Procope}
\email{cprocope@fb.com}
\affiliation{%
  \institution{Meta AI}
  \country{USA}
}
\author{Jeremy Sawruk$^{*}$}
\email{jsawruk@fb.com}
\affiliation{%
  \institution{Meta AI}
  \country{USA}
}
\author{Erin Wang}
\email{yulinw@fb.com}
\affiliation{%
  \institution{Meta AI}
  \country{USA}
}
\author{Polina Zvyagina}
\email{polinaz@fb.com}
\affiliation{%
  \institution{Meta AI}
  \country{USA}
}
\renewcommand{\shortauthors}{Adkins et al.}

\begin{abstract}
Specialized documentation techniques have been developed to communicate key facts about machine-learning (ML) systems and the datasets and models they rely on.
Techniques such as Datasheets, FactSheets, and Model Cards have taken a mainly descriptive approach, providing various details about the system components.
While the above information is essential for product developers and external experts to assess whether the ML system meets their requirements, other stakeholders might find it less actionable.
In particular, ML engineers need guidance on how to mitigate potential shortcomings in order to fix bugs or improve the system's performance.
We survey approaches that aim to provide such guidance in a prescriptive way.
We further propose a preliminary approach, called Method Cards, which aims to increase the transparency and reproducibility of ML systems by providing prescriptive documentation of commonly-used ML methods and techniques.
We showcase our proposal with an example in small object detection, and demonstrate
how Method Cards can communicate key considerations for model developers. We further highlight avenues for improving the user experience of ML engineers based on Method Cards.

\end{abstract}



\maketitle

\section{Introduction}
With the rapid adoption of machine learning (ML) in practical applications, ML-specific documentation has become crucial to their transparency and to the user experience of different stakeholders of ML.
In contrast to traditional software systems, the documentation of ML-based systems is an emerging area that poses new challenges and requirements.
A variety of initiatives have been undertaken over the past few years to address these challenges.
These initiatives aim to provide systematic ways to document ML-based systems.

The above-mentioned initiatives has focused on describing various components of ML-based systems.
For example, Datasheets for Datasets~\cite{Gebru2021Datasheets} focus on providing key details about the datasets used to develop ML models.
Similarly, Model Cards~\cite{mitchell2019model} aim to communicate key facts about individual models such as their intended use, training and evaluation data, and relevant metrics.
In addition, FactSheets~\cite{arnold2019factsheets} aim to act as declarations of conformity for AI services by providing relevant details to the consumers of these services.
While the above solutions advance the transparency of ML-based systems, their descriptive nature might limit their utility for certain stakeholders.
For example, ML engineers need specific information on how to retrain the system and how to mitigate certain problems, yet such knowledge is often sparsely documented.


We review and propose \textit{prescriptive} solutions that aim to provide greater transparency into ML-based systems and to improve the experience of their developers.
Our proposal caters mainly to expert stakeholders such as model developers and external model reviewers.
Our contributions encompass:
\begin{itemize}
    \item Motivating the need for a prescriptive approach to ML transparency, and surveying existing initiatives and forms of this approach (Section~\ref{sec:prescriptive}).
    \item Proposing Method Cards as a preliminary means to foster ML transparency and reproducibility, and improve the user experience of ML engineers (Section~\ref{sec:Method_Cards}).
\end{itemize}
In section~\ref{sec:discussion} we compare our proposals with previous work, and outline potential directions for future work.

\section{Background and Motivation}

Providing transparency into ML systems involves distinct challenges. 
Here we provide an overview of three complementary approaches that contribute to ML transparency along with a motivating example for prescriptive ones.
\vspace{-0.5mm}
\subsection{Transparency Through Documentation}
\label{sec:trans_thru_doc}
The ABOUT ML initiative~\cite{raji2019ml} advocates documentation as a practical intervention to provide clarity into decision making in ML systems for stakeholders.
The authors explain the value of both external and internal documentation such as establishing trust and demonstrating fairness~\cite{holstein2019improving}.
Furthermore, the authors argue that documentation is both an artifact and a process, showing how developing the documentation fosters ML developers to think critically about every step in the ML lifecycle. Accordingly, the initiative is led by the Partnership on AI consortium which continues to develop this process.

Documenting software systems is a long-standing goal of software engineering.
Some of the emerging documentation initiatives for ML-based systems have adapted existing methods while others necessarily take a novel approach.
Datasheets~\cite{Gebru2021Datasheets}, and to an extent Model Cards~\cite{mitchell2019model}, grew from experiences with hardware specification documentation. Dataset Nutrition Labels \cite{holland2018dataset, chmielinski2020dataset}  and FactSheets~\cite{arnold2019factsheets} are further prominent examples of these initiatives.
Hind et al~\cite{hind2020experiences} report on the experience of AI teams in using FactSheets, and provide recommendations for easing the collection and flexible presentation of AI facts to promote transparency.


\vspace{-0.5mm}
\subsection{Transparency Through Reproducibility}

The ability of stakeholders to reproduce an ML system significantly increases its transparency~\cite{haibe2020transparency}.
In addition to testing the system on their own data points, reproducibility enables stakeholders to retrain the ML models independently.
Besides ensuring trust in the system and its components, the ability to retrain the ML models is crucial to analyze and debug these models.

Reproducibility in ML cannot be simply solved by making the source code available. 
The nature of ML algorithms and how they are trained make it challenging to warrant \textit{reproducibility of the results}~\cite{sculley2015hidden}.
Bell and Kampman~\cite{bell2021perspectives} proposed ideas to foster this reproducibility, inspired by the recent reformation in psychology.
Examples for these are Multiverse analysis~\cite{steegen2016increasing}, preregistration~\cite{nosek2012scientific}, and encouraging the publication of negative results.
Likewise, several proposals have been made to quantify and accordingly improve reproducibility~\cite{raff2019step, pineau2021improving}.

\vspace{-0.5mm}
\subsection{Transparency Through Interpretability}

Lipton~\cite{lipton2018mythos} analyzes in detail the notion of transparency in interpretability.
The author argues how transparency in this context is the opposite of ``black-boxness'' as it helps understanding the mechanism by which the model works.
Such understanding can be at the level of the entire model, at the level of individual components such as parameters, and at the level of the training algorithm.

Weller~\cite{weller2019transparency} identifies different use cases of transparency, and distinguishes between two types of interpretability solutions to support them: global (how an overall system works) and local (explaining a particular prediction).
The author surveys available techniques for both types and argues how the transparency enabled by them can be provide insights into important model characteristics such as robustness and fairness.

\subsection{Illustrative Example}
\label{sec:illustrative}
Consider a team of ML engineers tasked with developing an image object detector for a specific application.
The team has a variety of choices to make such as the model type to use (e.g. convolutional networks vs. Vision Transformers), the training paradigm (e.g. supervised vs. self-supervised learning, etc.), the architecture type (e.g. ResNet vs. VGGNet), the optimizer to use, and the input preprocessing operations to mention a few. To speed up the ideation process and to obtain a benchmarking baseline, the team starts by finetuning a Single-Shot Detector (SSD) with a ResNet-50 as a backbone convolutional network pre-trained on ImageNet. This transfer-learning paradigm is very popular, especially when the training data is limited. They resize their images to $640 x 480$ to maintain the aspect ratio, and train their model following the same training script used to train the backbone model.
However, there are a variety of nuances that impact the performance and reliability of their model. A few examples of these issues are:
\begin{itemize}
    \item The resizing operation in popular deep-learning frameworks has a subtle aliasing flaw \cite{parmar2021buggy}.
    \item The input size used was shown to induce parity issues in ResNet models~\cite{alsallakh2021debugging}, leading to skewness in the learned filters.
    \item The default 0-padding used in ResNet was shown to impact small object detection~\cite{mindThePad}, leading to blind spots.
    \item Transfer learning from a supervised model trained on ImageNet can lead to conflicts between the ImageNet categories and the objects of the target domain~\cite{zoph2020rethinking}.
\end{itemize}

Fortunately, it is possible to mitigate the above issues by making careful architectural, training, and preprocessing choices.
However, these issues might not be widely known among practitioners and are often left unaddressed.
Our goal is to provide ML engineers with actionable guidance on available ML methods and how to use them effectively in the systems they intend to develop.
In contrast to the descriptive documentation techniques mentioned in Section~\ref{sec:trans_thru_doc}, our guidance follows a prescriptive approach. 
As a rough metaphor, Model Cards and FactSheets are analogous to user manuals that are shipped with software programs and automobiles.
While we aim to create prescriptive solutions that are analogous to computer programming recipes~\cite{press1989numerical} and car repair manuals.

\section{A Prescriptive Approach to ML Transparency}
\label{sec:prescriptive}
Software engineering has a long history of employing prescriptive models \cite{ludewig2003models} for various purposes such as effective diffusion \cite{raghavan1989diffusing}, safety certification \cite{hawkins2013assurance}, and error analysis \cite{meng2019new}.
A key distinction between descriptive and prescriptive models is that the former describe an existing system, while the latter specify how a new system should be created~\cite{ludewig2003models}.

\begin{figure*}[th!]
 \centering
 \includegraphics[width=0.9\linewidth]{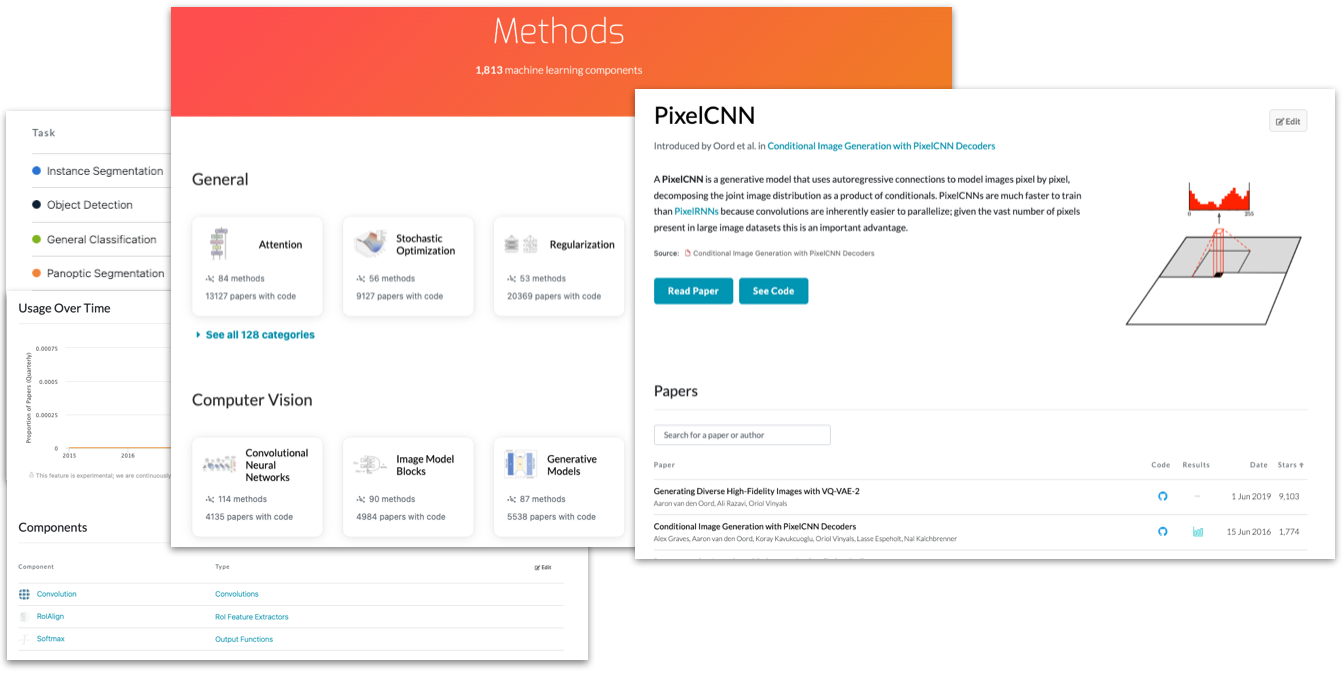}
 \caption {
 Documenting ML methods in Papers With Code. For each method, the documentation includes brief textual and graphical descriptions, a list of papers that introduce or use the method, and a list of ML tasks the method is suited for.}
 \label{fig:PWC} 
\end{figure*}

Descriptive approaches such as documenting general information of an existing system can sometimes be viewed by ML practitioners as time-consuming, lacking clear incentives, and/or as having intelligibility concerns \cite{miceli2021documenting}. 
Prescriptive approaches, on the other hand, provide explicit instructions on how to develop and deploy a solution or how to handle unexpected situations. This is useful in a variety of engineering disciplines both for modeling  solutions \cite{finger1989review,heldal2016descriptive}  and for guiding troubleshooting efforts \cite{eliasson2015architecting, wong2019prescriptive}. Specific methods have been devised to support the documentation process of prescriptive software models \cite{clements2003documenting}.

The Machine Learning community has recently started employing prescriptive approaches to developing ML systems.
The prescriptive nature of such approaches directly contribute to the transparency of the ML system, especially from the perspective of ML engineers and external reviewers.
The following initiatives by the ML community involve prescriptive aspects and are targeted either at ML engineers or at ML researchers.

\paragraph{\textbf{Guidelines and Design Patterns}}
A number of design patterns have recently emerged in the ML community, both among practitioners~\cite{Apple2019Guidelines, Google2019Guidebook, lakshmanan2020machine, Microsoft2021HAX, Shibui2020patterns}, and among researchers~\cite{washizaki2022software, yokoyama2019machine}.
These patterns tackle various aspects of model development and deployment, data and problem representation, training, serving and operation, reproducibility, quality-control and responsible AI usage.
These patterns as well as complementary anti-patterns~\cite{washizaki2019studying} capture best practices and are very helpful to replicate successful solutions in new applications.

\paragraph{\textbf{Recipes}}
    As one of the leading deep-learning frameworks, PyTorch~\cite{Paszke2019PyTorch} maintains a list of recipes~\footnote{Official PyTorch Recipes: \url{https://pytorch.org/tutorials/recipes/recipes_index.html}}, defined as 
    ``bite-sized, actionable examples of how to use specific PyTorch features''.
    In contrast to the standard documentation~\footnote{PyTorch Documentation \url{https://pytorch.org/docs/stable/index.html}}, these recipes focus on executable solutions and demonstrate best practices established by the framework developers and its community.
    
\paragraph{\textbf{PWC Methods}}
The Papers With Code (PWC) community project has recently started maintaining a catalog of ML methods proposed in the research community for various ML tasks~\footnote{\url{https://paperswithcode.com/methods}}.
The catalog features a categorization of the method by task or data modality as well as an informative community-generated description of each method, along with statistics about its usage (Figure~\ref{fig:PWC}).
Unlike Model Cards and FactSheets, these descriptions are independent of any specific models or systems, focusing instead on the underlying ML methods available in the literature, with ML researchers as the main audience. 

\paragraph{\textbf{Reproducibility Checklists}}
Over the past few years, the ML research community has developed a reproducibility checklist, with the goal of ensuring that research articles and results are easy to reproduce~\cite{pineau2021improving}.
The checklist includes specific instructions regarding the models and algorithms presented, the datasets used, and the code shared, in addition to expectations regarding any theoretical claim or experimental results.

\paragraph{\textbf{ML Cheat Sheets}}
A variety of cheat sheets have been developed to provide an overview of ML algorithms.
These sheets typically provide a flow chart on how to choose an algorithm for a specific problem, as in the ones developed by Microsoft Azure~\footnote{\url{https://blogs.sas.com/content/subconsciousmusings/2020/12/09/machine-learning-algorithm-use/}} or by SAS~\footnote{\url{https://docs.microsoft.com/en-us/azure/machine-learning/algorithm-cheat-sheet}}. Similarity, ML Flashcards serve as a quick reference of ML concepts~\footnote{\url{https://machinelearningflashcards.com/}}.
Additionally, several practitioners have compiled best practices for training ML models~\cite{ang2018bunch, he2019bag}.

\paragraph{\textbf{Interactive Hyperparameter Visualizations}}
A variety of interactive widgets aim to help ML engineers in making informed choices of their model's hyperparameters by visualizing how these choices impact the model.
Examples include basic convolution arithmetic~\footnote{Convolution Visualizer by Edward Yang: \url{ https://ezyang.github.io/convolution-visualizer/}}, choosing t-SNE parameters~\cite{wattenberg2016use}, and understanding the impact of padding choices~\cite{yuan2021convolution}. 



\section{Method Cards}
\label{sec:Method_Cards}
We propose Method Cards to guide ML engineers throughout the process of model development.
Analogous to Model Cards, these cards aim to communicate key information about  ML methods. The information comprises both prescriptive and descriptive elements, putting the main focus on ensuring that ML engineers are able to use these methods properly. 
The cards aim to support developers at multiple stages of the model-development process such as training, testing, and debugging. For this purpose, we propose a structure for the cards, outlined in Figure~\ref{fig:method_card}.

For each prompt in  Method Cards, the method creators are expected to provide sufficient instructions and documentation to guide ML engineers.
These instructions should help the engineers in choosing suited preprocessing steps, model components, and hyperparameters for their task.
Furthermore, the instructions should make these engineers aware of how their choices potentially impact the model's behavior, how to evaluate this impact, and how to handle prediction errors accordingly.
 Finally, the instructions should explicitly inform the engineers on responsible usage of the method by addressing potential fairness and privacy concerns.

The instructions and documentation needed for the prompts in Method Cards can be enriched with the prescriptive approaches outlined in Section~\ref{sec:prescriptive}.
IBM's FactSheets provide several examples that demonstrate how to format and deliver such rich  information at multiple levels of detail in a usable way.

\subsection{Granularity of ML Methods}
The sections and prompts we propose in Figure~\ref{fig:method_card} focus on ML methods that are sufficient to produce a proper ML model with defined input, output, and task. Examples for these are object detection methods such as Single-shot Detectors~\cite{liu2016ssdV1} and language modelling methods such as Generative Pre-trained Transformers (GPT)~\cite{radford2019language}. It is possible to create Model Cards for the models created using these methods.


\setlength{\fboxsep}{1em}
\begin{figure*}[th!]
\begin{center}
    \textbf{Method Card Template}
\end{center}
\begin{mdframed}
\begin{multicols}{2}

\textbf{Basic Method Information}
\begin{itemize}
\item Name, version, and application domain(s).
\item Method purpose and appropriate uses.
\item Method definition, published literature, reference implementation.
\item Example input and output.
\end{itemize}

\textbf{Safety and Troubleshooting}
\begin{itemize}
\item Inappropriate uses and common usage pitfalls.
\item Known weaknesses, biases, and privacy leakage.
\item How to detect biases in the model internals.
\item Common failure modes, potential root causes, and possible mitigations via hyperparameter tuning or training data expansion.
\end{itemize}

\textbf{Data Preparation}
\begin{itemize}
\item Input and output format, shape, and data type.
\item Data transformation and normalization.
\item Recommended sampling and balancing.
\item Recommended batching scheme and batch size.
\item Required data augmentation and shuffling.
\item Validation and train-test splitting schemes.
\end{itemize}

\textbf{Modelling and Optimization}
\begin{itemize}
\item Architecture family and components used.
\item A list of hyperparameters, along with applicable values and their known impact.
\item Training objective(s), loss(es), and optimizer(s).
\item Parameter initialization / self pre-training / transfer from a trained baseline (specify datasets).
\item Regularization scheme, capacity selection.
\item If applicable, learning rate and schedulers.
\item Weight quantization, recommended bit depth.
\item Possibilities to compile the model graph.
\item Parallelization at training and inference time.
\item Recommended model compression techniques.
\end{itemize}

\textbf{Method Benchmarking}
\begin{itemize}
\item Performance metric(s) and applicable threshold(s).
\item Threshold selection.
\item Fairness evaluation and subgroup comparison.
\item Overfitting detection.
\item Training and inference time efficiency.
\item Available benchmarks.
\end{itemize}

\textbf{Interpretability and Explainability}
\begin{itemize}
\item Applicable feature attribution methods, and how they can help explain model predictions.
\item How to identify influential training instances behind a specific model prediction.
\item How to identify internal concepts and features learned using the method.
\end{itemize}

\textbf{Robustness}
\begin{itemize}
\item Known vulnerabilities to adversarial attacks, and recommended mitigation.
\item Out-of-distribution behavior.
\item Detecting and mitigating data and model drifts.
\end{itemize}

\end{multicols}
\end{mdframed}

 \caption {
 Suggested sections in Method Cards, along with the prompts they capture.
 Providing instructions for each prompt offers guidance to ML engineers and helps external reviewers evaluate the maturity of the model development processes.}.
 \label{fig:method_card} 
\end{figure*}

We refer to ML methods that focus on specific parts of a model as \textbf{ML components}, following the  PWC terminology. Examples of these are pooling, regularization, and normalization operations.
These components correspond to specific prompts in our proposed Method Cards, specifically, the ones in ``Data Preparation'' and ``Modelling and Training''.
The ML community is constantly studying these components, identifying their strengths and weaknesses in different ML applications.
PWC provides a good overview of these studies, which can help the creators of Method Cards provide actionable recommendations about these components.

Likewise, \textbf{ML systems} often encompass multiple ML models, as well as non-ML components such as data acquisition and human-in-the-loop interfaces. Examples of these are fraud detection systems that rely on ML models to identify abnormal behavior and on human reviewers to take appropriate action when an automated decision could not be determined reliably.
Providing transparency into such systems and services require artifacts similar to FactSheets.

\subsection{Maintaining Method Cards}

Within corporations, ML teams can benefit from maintaining method cards for their common use cases. 
These cards can be curated collectively and updated regularly as the team experiments with various components and parameterizations, and as they gain insights about their method's weaknesses and edge cases.
Such documentation is important for onboarding new ML engineers and to enable them to maintain the ML models and to retrain them when needed.
As such, Method Cards offer a systematic way to communicate best ML practices identified by the team.

Since ML methods share a variety of components and can have similar goals, their cards can naturally overlap.
It is often possible to categorize these methods into a hierarchy, and to document shared information into cards that correspond to high levels in the hierarchy. For example, methods based on convolutional neural networks can inherit generic considerations related to these models such as choosing a suited input size~(Section~\ref{sec:illustrative}).
Likewise, object detection methods can inherent common considerations such as evaluation metrics and data preparation.
The hierarchy used to categorize PWC Methods could serve as a useful reference.


\subsection{An Example of Method Cards}
In the appendix we provide an example method card, created for traffic light detection using an SSD.
The card exemplifies how a team of ML engineers can collectively document the nuances mentioned in Section~\ref{sec:illustrative}, both based on published literature and on their own observations and experiences.
Notice how the information presented is independent of a specific model.
For example, the challenges of using SSDs for small object detection are inherent to the method.
Likewise, many challenges arise from the nature of the problem, such as the similarity of red traffic lights with tail lights.

In the example, we used concise descriptions and instructions to avoid turning the curation and consumption of Method Cards into a burden.
Nevertheless, the curators can elaborate on certain prompts by providing links to executable notebooks, articles, and further materials needed to clarify the prompts.



\section{Discussion}

\label{sec:discussion}

\paragraph{Possible Applications}
Besides serving as a tool for prescriptive documentation, Method Cards, can further support building interactive tools to improve ML development processes. 
For example, a formal specification of certain prompts can be used to allow generating ML code templates.
These templates serve as a useful starting points that ML developers can adapt, and help reducing potential coding errors as demonstrated in ClassyVision templates~\cite{adcock2019classy}.

Our effort to explore prescriptive solutions started in response to preliminary interviews we conducted with ML engineers who work on various ML problems in practical applications. Of their frequent pain points are the ability to make informed choices when designing and training new models, as well as finding solutions to improve the accuracy of existing models. 
As such, Method Cards aim to fill one of the gaps identified by the Model Cards authors:
\begin{quote}
\textit{It seems unlikely, at least in the near term, that model cards could be standardized or formalized to a degree needed to prevent misleading representations of model results (whether intended or unintended). It is therefore important to consider model cards as one transparency tool among many, which could include, for example, algorithmic auditing by third-parties [...]'' \hspace{2mm}  - \cite[p. 9]{mitchell2019model}.} 
\end{quote}
While Model Cards and FactSheets put main focus on documenting existing models, Method Cards focus more on the underlying methodical and algorithmic choices that need to be considered when creating and training these models. As a rough analogy, if Model Cards and FactSheets provide nutrition information about cooked meals, Method Cards provide the recipes.

\paragraph{Comparison with PWC Methods} PWC Methods have mainly focused on offering a catalog of methods extracted from ML literature, designed to be usable mainly for researchers.
On the other hand, Method Cards have ML engineers and algorithmic reviewers as target users, focusing on transparent, reproducible, and responsible usage of these methods.
Ultimately, interested stakeholders in the ML community can augment PWC Methods with the prescriptive information we proposed in our template to help meet these requirements.

\paragraph{The Usability of Method Cards}
A major challenge in offering documentation-based transparency is the effort needed by ML engineers to develop the needed artifacts~\cite{miceli2021documenting}, compared with the perceived benefits.
Finding the appropriate level of detail is a key to alleviate the burden and maximize the value, as the authors of FactSheets reported~\cite{hind2020experiences}.
Accordingly, we consider the template suggested in Figure~\ref{fig:method_card} to be an reference example, and encourage ML engineers to refine the prompts to fit their needs.
Furthermore, interactive interfaces can help presenting the information at multiple levels of detail. For example, detailed answers can help onboard new team members, while summarized bullet lists are more suited for external reviewers.



\section{Conclusion}
We presented an overview of prescriptive approaches in ML documentation and their ability to facilitate transparency into ML algorithms.
We further presented Method Cards that leverage descriptive and prescriptive elements to directly address the challenges surrounding algorithmic transparency.
Unlike Model Cards, these cards do not describe specific ML models, focusing instead on providing guidance on how to properly use ML methods to define and train these models.
This helps increase reproducibility of ML and enables external reviewers to determine the appropriateness of the methods used in ML-based solutions.
Furthermore, this enables ML engineers to develop best practices to mitigate potential shortcomings and to improve the system's performance.
We presented an example of Method Cards along with potential applications in generating templates and powering static analyzers of ML code. 


\bibliographystyle{ACM-Reference-Format}
\bibliography{main}


\begin{thebibliography}{50}


\ifx \showCODEN    \undefined \def \showCODEN     #1{\unskip}     \fi
\ifx \showDOI      \undefined \def \showDOI       #1{#1}\fi
\ifx \showISBNx    \undefined \def \showISBNx     #1{\unskip}     \fi
\ifx \showISBNxiii \undefined \def \showISBNxiii  #1{\unskip}     \fi
\ifx \showISSN     \undefined \def \showISSN      #1{\unskip}     \fi
\ifx \showLCCN     \undefined \def \showLCCN      #1{\unskip}     \fi
\ifx \shownote     \undefined \def \shownote      #1{#1}          \fi
\ifx \showarticletitle \undefined \def \showarticletitle #1{#1}   \fi
\ifx \showURL      \undefined \def \showURL       {\relax}        \fi
\providecommand\bibfield[2]{#2}
\providecommand\bibinfo[2]{#2}
\providecommand\natexlab[1]{#1}
\providecommand\showeprint[2][]{arXiv:#2}

\bibitem[\protect\citeauthoryear{{Adcock}, {Reis}, {Singh}, {Yan}, {van der
  Maaten}, {Zhang}, {Motwani}, {Guerin}, {Goyal}, {Misra}, {Gustafson},
  {Changhan}, and {Goyal}}{{Adcock} et~al\mbox{.}}{2019}]%
        {adcock2019classy}
\bibfield{author}{\bibinfo{person}{A. {Adcock}}, \bibinfo{person}{V. {Reis}},
  \bibinfo{person}{M. {Singh}}, \bibinfo{person}{Z. {Yan}}, \bibinfo{person}{L.
  {van der Maaten}}, \bibinfo{person}{K. {Zhang}}, \bibinfo{person}{S.
  {Motwani}}, \bibinfo{person}{J. {Guerin}}, \bibinfo{person}{N. {Goyal}},
  \bibinfo{person}{I. {Misra}}, \bibinfo{person}{L. {Gustafson}},
  \bibinfo{person}{C. {Changhan}}, {and} \bibinfo{person}{P. {Goyal}}.}
  \bibinfo{year}{2019}\natexlab{}.
\newblock \bibinfo{title}{Classy Vision}.
\newblock
  \bibinfo{howpublished}{\url{https://github.com/facebookresearch/ClassyVision}}.
\newblock


\bibitem[\protect\citeauthoryear{Alsallakh, Kokhlikyan, Miglani, Muttepawar,
  Wang, Zhang, Adkins, and Reblitz-Richardson}{Alsallakh
  et~al\mbox{.}}{2021a}]%
        {alsallakh2021debugging}
\bibfield{author}{\bibinfo{person}{B. Alsallakh}, \bibinfo{person}{N.
  Kokhlikyan}, \bibinfo{person}{V. Miglani}, \bibinfo{person}{S. Muttepawar},
  \bibinfo{person}{E. Wang}, \bibinfo{person}{S. Zhang}, \bibinfo{person}{D.
  Adkins}, {and} \bibinfo{person}{O. Reblitz-Richardson}.}
  \bibinfo{year}{2021}\natexlab{a}.
\newblock \showarticletitle{Debugging the Internals of Convolutional Networks}.
  In \bibinfo{booktitle}{\emph{eXplainable AI approaches for debugging and
  diagnosis - NeurIPS Workshop.}}
\newblock


\bibitem[\protect\citeauthoryear{Alsallakh, Kokhlikyan, Miglani, Yuan, and
  Reblitz-Richardson}{Alsallakh et~al\mbox{.}}{2021b}]%
        {mindThePad}
\bibfield{author}{\bibinfo{person}{B. Alsallakh}, \bibinfo{person}{N.
  Kokhlikyan}, \bibinfo{person}{V. Miglani}, \bibinfo{person}{J. Yuan}, {and}
  \bibinfo{person}{O. Reblitz-Richardson}.} \bibinfo{year}{2021}\natexlab{b}.
\newblock \showarticletitle{Mind the Pad -- CNNs Can Develop Blind Spots}. In
  \bibinfo{booktitle}{\emph{Intl. Conference on Learning Representations
  (ICLR)}}.
\newblock


\bibitem[\protect\citeauthoryear{Ang}{Ang}{2018}]%
        {ang2018bunch}
\bibfield{author}{\bibinfo{person}{Long Ang}.} \bibinfo{year}{2018}\natexlab{}.
\newblock \showarticletitle{A bunch of tips and tricks for training deep neural
  networks}.
\newblock  (\bibinfo{year}{2018}).
\newblock
\urldef\tempurl%
\url{https://towardsdatascience.com/a-bunch-of-tips-and-tricks-for-training-deep-neural-networks-3ca24c31ddc8}
\showURL{%
\tempurl}


\bibitem[\protect\citeauthoryear{Apple}{Apple}{2019}]%
        {Apple2019Guidelines}
\bibfield{author}{\bibinfo{person}{Apple}.} \bibinfo{year}{2019}\natexlab{}.
\newblock \showarticletitle{Human interface guidelines for machine learning}.
\newblock  (\bibinfo{year}{2019}).
\newblock
\urldef\tempurl%
\url{https://developer.apple.com/design/human-interface-guidelines/machine-learning/}
\showURL{%
\tempurl}


\bibitem[\protect\citeauthoryear{Arnold, Bellamy, Hind, Houde, Mehta,
  Mojsilovi{\'c}, Nair, Ramamurthy, Olteanu, Piorkowski, et~al\mbox{.}}{Arnold
  et~al\mbox{.}}{2019}]%
        {arnold2019factsheets}
\bibfield{author}{\bibinfo{person}{Matthew Arnold}, \bibinfo{person}{Rachel~KE
  Bellamy}, \bibinfo{person}{Michael Hind}, \bibinfo{person}{Stephanie Houde},
  \bibinfo{person}{Sameep Mehta}, \bibinfo{person}{Aleksandra Mojsilovi{\'c}},
  \bibinfo{person}{Ravi Nair}, \bibinfo{person}{K~Natesan Ramamurthy},
  \bibinfo{person}{Alexandra Olteanu}, \bibinfo{person}{David Piorkowski},
  {et~al\mbox{.}}} \bibinfo{year}{2019}\natexlab{}.
\newblock \showarticletitle{FactSheets: Increasing trust in AI services through
  supplier's declarations of conformity}.
\newblock \bibinfo{journal}{\emph{IBM Journal of Research and Development}}
  \bibinfo{volume}{63}, \bibinfo{number}{4/5} (\bibinfo{year}{2019}),
  \bibinfo{pages}{6--1}.
\newblock


\bibitem[\protect\citeauthoryear{Behrendt, Novak, and Botros}{Behrendt
  et~al\mbox{.}}{2017}]%
        {behrendt2017deep}
\bibfield{author}{\bibinfo{person}{Karsten Behrendt}, \bibinfo{person}{Libor
  Novak}, {and} \bibinfo{person}{Rami Botros}.}
  \bibinfo{year}{2017}\natexlab{}.
\newblock \showarticletitle{A deep learning approach to traffic lights:
  Detection, tracking, and classification}. In \bibinfo{booktitle}{\emph{2017
  IEEE International Conference on Robotics and Automation (ICRA)}}. IEEE,
  \bibinfo{pages}{1370--1377}.
\newblock


\bibitem[\protect\citeauthoryear{Bell and Kampman}{Bell and Kampman}{2021}]%
        {bell2021perspectives}
\bibfield{author}{\bibinfo{person}{Samuel~J Bell} {and} \bibinfo{person}{Onno~P
  Kampman}.} \bibinfo{year}{2021}\natexlab{}.
\newblock \showarticletitle{Perspectives on Machine Learning from Psychology's
  Reproducibility Crisis}.
\newblock \bibinfo{journal}{\emph{ICLR Workshop on Science and Engineering of
  Deep Learning arXiv:2104.08878}} (\bibinfo{year}{2021}).
\newblock


\bibitem[\protect\citeauthoryear{Chmielinski, Newman, Taylor, Joseph, Thomas,
  Yurkofsky, and Qiu}{Chmielinski et~al\mbox{.}}{2020}]%
        {chmielinski2020dataset}
\bibfield{author}{\bibinfo{person}{Kasia~S Chmielinski}, \bibinfo{person}{Sarah
  Newman}, \bibinfo{person}{Matt Taylor}, \bibinfo{person}{Josh Joseph},
  \bibinfo{person}{Kemi Thomas}, \bibinfo{person}{Jessica Yurkofsky}, {and}
  \bibinfo{person}{Yue~Chelsea Qiu}.} \bibinfo{year}{2020}\natexlab{}.
\newblock \showarticletitle{The dataset nutrition label (2nd Gen): Leveraging
  context to mitigate harms in artificial intelligence}. In
  \bibinfo{booktitle}{\emph{NeurIPS Workshop on Dataset Curation and
  Security}}.
\newblock


\bibitem[\protect\citeauthoryear{Clements, Garlan, Little, Nord, and
  Stafford}{Clements et~al\mbox{.}}{2003}]%
        {clements2003documenting}
\bibfield{author}{\bibinfo{person}{Paul Clements}, \bibinfo{person}{David
  Garlan}, \bibinfo{person}{Reed Little}, \bibinfo{person}{Robert Nord}, {and}
  \bibinfo{person}{Judith Stafford}.} \bibinfo{year}{2003}\natexlab{}.
\newblock \showarticletitle{Documenting software architectures: views and
  beyond}. In \bibinfo{booktitle}{\emph{25th International Conference on
  Software Engineering, 2003. Proceedings.}} IEEE, \bibinfo{pages}{740--741}.
\newblock


\bibitem[\protect\citeauthoryear{Eliasson, Heldal, Pelliccione, and
  Lantz}{Eliasson et~al\mbox{.}}{2015}]%
        {eliasson2015architecting}
\bibfield{author}{\bibinfo{person}{Ulf Eliasson}, \bibinfo{person}{Rogardt
  Heldal}, \bibinfo{person}{Patrizio Pelliccione}, {and} \bibinfo{person}{Jonn
  Lantz}.} \bibinfo{year}{2015}\natexlab{}.
\newblock \showarticletitle{Architecting in the automotive domain: Descriptive
  vs prescriptive architecture}. In \bibinfo{booktitle}{\emph{2015 12th Working
  IEEE/IFIP Conference on Software Architecture}}. IEEE,
  \bibinfo{pages}{115--118}.
\newblock


\bibitem[\protect\citeauthoryear{Finger and Dixon}{Finger and Dixon}{1989}]%
        {finger1989review}
\bibfield{author}{\bibinfo{person}{Susan Finger} {and} \bibinfo{person}{John~R
  Dixon}.} \bibinfo{year}{1989}\natexlab{}.
\newblock \showarticletitle{A review of research in mechanical engineering
  design. Part I: Descriptive, prescriptive, and computer-based models of
  design processes}.
\newblock \bibinfo{journal}{\emph{Research in engineering design}}
  \bibinfo{volume}{1}, \bibinfo{number}{1} (\bibinfo{year}{1989}),
  \bibinfo{pages}{51--67}.
\newblock


\bibitem[\protect\citeauthoryear{Gebru, Morgenstern, Vecchione, Vaughan,
  Wallach, III, and Crawford}{Gebru et~al\mbox{.}}{2021}]%
        {Gebru2021Datasheets}
\bibfield{author}{\bibinfo{person}{T. Gebru}, \bibinfo{person}{J. Morgenstern},
  \bibinfo{person}{B. Vecchione}, \bibinfo{person}{J.~Wortman Vaughan},
  \bibinfo{person}{H. Wallach}, \bibinfo{person}{Hal~Daum\'{e} III}, {and}
  \bibinfo{person}{K. Crawford}.} \bibinfo{year}{2021}\natexlab{}.
\newblock \showarticletitle{Datasheets for Datasets}.
\newblock \bibinfo{journal}{\emph{Commun. ACM}} \bibinfo{volume}{64},
  \bibinfo{number}{12} (\bibinfo{date}{nov} \bibinfo{year}{2021}),
  \bibinfo{pages}{86–92}.
\newblock
\showISSN{0001-0782}
\urldef\tempurl%
\url{https://doi.org/10.1145/3458723}
\showDOI{\tempurl}


\bibitem[\protect\citeauthoryear{Geirhos, Rubisch, Michaelis, Bethge, Wichmann,
  and Brendel}{Geirhos et~al\mbox{.}}{2019}]%
        {geirhos2018imagenettrained}
\bibfield{author}{\bibinfo{person}{Robert Geirhos}, \bibinfo{person}{Patricia
  Rubisch}, \bibinfo{person}{Claudio Michaelis}, \bibinfo{person}{Matthias
  Bethge}, \bibinfo{person}{Felix~A. Wichmann}, {and} \bibinfo{person}{Wieland
  Brendel}.} \bibinfo{year}{2019}\natexlab{}.
\newblock \showarticletitle{ImageNet-trained {CNN}s are biased towards texture;
  increasing shape bias improves accuracy and robustness.}. In
  \bibinfo{booktitle}{\emph{International Conference on Learning
  Representations}}.
\newblock
\urldef\tempurl%
\url{https://openreview.net/forum?id=Bygh9j09KX}
\showURL{%
\tempurl}


\bibitem[\protect\citeauthoryear{Haibe-Kains, Adam, Hosny, Khodakarami,
  Waldron, Wang, McIntosh, Goldenberg, Kundaje, Greene,
  et~al\mbox{.}}{Haibe-Kains et~al\mbox{.}}{2020}]%
        {haibe2020transparency}
\bibfield{author}{\bibinfo{person}{Benjamin Haibe-Kains},
  \bibinfo{person}{George~Alexandru Adam}, \bibinfo{person}{Ahmed Hosny},
  \bibinfo{person}{Farnoosh Khodakarami}, \bibinfo{person}{Levi Waldron},
  \bibinfo{person}{Bo Wang}, \bibinfo{person}{Chris McIntosh},
  \bibinfo{person}{Anna Goldenberg}, \bibinfo{person}{Anshul Kundaje},
  \bibinfo{person}{Casey~S Greene}, {et~al\mbox{.}}}
  \bibinfo{year}{2020}\natexlab{}.
\newblock \showarticletitle{Transparency and reproducibility in artificial
  intelligence}.
\newblock \bibinfo{journal}{\emph{Nature}} \bibinfo{volume}{586},
  \bibinfo{number}{7829} (\bibinfo{year}{2020}), \bibinfo{pages}{E14--E16}.
\newblock


\bibitem[\protect\citeauthoryear{Hawkins, Habli, Kelly, and McDermid}{Hawkins
  et~al\mbox{.}}{2013}]%
        {hawkins2013assurance}
\bibfield{author}{\bibinfo{person}{Richard Hawkins}, \bibinfo{person}{Ibrahim
  Habli}, \bibinfo{person}{Tim Kelly}, {and} \bibinfo{person}{John McDermid}.}
  \bibinfo{year}{2013}\natexlab{}.
\newblock \showarticletitle{Assurance cases and prescriptive software safety
  certification: A comparative study}.
\newblock \bibinfo{journal}{\emph{Safety science}}  \bibinfo{volume}{59}
  (\bibinfo{year}{2013}), \bibinfo{pages}{55--71}.
\newblock


\bibitem[\protect\citeauthoryear{He, Zhang, Zhang, Zhang, Xie, and Li}{He
  et~al\mbox{.}}{2019}]%
        {he2019bag}
\bibfield{author}{\bibinfo{person}{T. He}, \bibinfo{person}{Z. Zhang},
  \bibinfo{person}{H. Zhang}, \bibinfo{person}{Z. Zhang}, \bibinfo{person}{J.
  Xie}, {and} \bibinfo{person}{M. Li}.} \bibinfo{year}{2019}\natexlab{}.
\newblock \showarticletitle{Bag of tricks for image classification with
  convolutional neural networks}. In \bibinfo{booktitle}{\emph{Proceedings of
  the IEEE/CVF Conference on Computer Vision and Pattern Recognition}}.
  \bibinfo{pages}{558--567}.
\newblock


\bibitem[\protect\citeauthoryear{Heldal, Pelliccione, Eliasson, Lantz, Derehag,
  and Whittle}{Heldal et~al\mbox{.}}{2016}]%
        {heldal2016descriptive}
\bibfield{author}{\bibinfo{person}{Rogardt Heldal}, \bibinfo{person}{Patrizio
  Pelliccione}, \bibinfo{person}{Ulf Eliasson}, \bibinfo{person}{Jonn Lantz},
  \bibinfo{person}{Jesper Derehag}, {and} \bibinfo{person}{Jon Whittle}.}
  \bibinfo{year}{2016}\natexlab{}.
\newblock \showarticletitle{Descriptive vs prescriptive models in industry}. In
  \bibinfo{booktitle}{\emph{Proceedings of the acm/ieee 19th international
  conference on model driven engineering languages and systems}}.
  \bibinfo{pages}{216--226}.
\newblock


\bibitem[\protect\citeauthoryear{Hind, Houde, Martino, Mojsilovic, Piorkowski,
  Richards, and Varshney}{Hind et~al\mbox{.}}{2020}]%
        {hind2020experiences}
\bibfield{author}{\bibinfo{person}{Michael Hind}, \bibinfo{person}{Stephanie
  Houde}, \bibinfo{person}{Jacquelyn Martino}, \bibinfo{person}{Aleksandra
  Mojsilovic}, \bibinfo{person}{David Piorkowski}, \bibinfo{person}{John
  Richards}, {and} \bibinfo{person}{Kush~R Varshney}.}
  \bibinfo{year}{2020}\natexlab{}.
\newblock \showarticletitle{Experiences with improving the transparency of {AI}
  models and services}. In \bibinfo{booktitle}{\emph{Extended Abstracts of the
  CHI Conference on Human Factors in Computing Systems}}.
  \bibinfo{pages}{1--8}.
\newblock


\bibitem[\protect\citeauthoryear{Holland, Hosny, Newman, Joseph, and
  Chmielinski}{Holland et~al\mbox{.}}{2018}]%
        {holland2018dataset}
\bibfield{author}{\bibinfo{person}{Sarah Holland}, \bibinfo{person}{Ahmed
  Hosny}, \bibinfo{person}{Sarah Newman}, \bibinfo{person}{Joshua Joseph},
  {and} \bibinfo{person}{Kasia Chmielinski}.} \bibinfo{year}{2018}\natexlab{}.
\newblock \showarticletitle{The dataset nutrition label: A framework to drive
  higher data quality standards}.
\newblock \bibinfo{journal}{\emph{arXiv preprint arXiv:1805.03677}}
  (\bibinfo{year}{2018}).
\newblock


\bibitem[\protect\citeauthoryear{Holstein, Wortman~Vaughan, Daum{\'e}~III,
  Dudik, and Wallach}{Holstein et~al\mbox{.}}{2019}]%
        {holstein2019improving}
\bibfield{author}{\bibinfo{person}{Kenneth Holstein}, \bibinfo{person}{Jennifer
  Wortman~Vaughan}, \bibinfo{person}{Hal Daum{\'e}~III}, \bibinfo{person}{Miro
  Dudik}, {and} \bibinfo{person}{Hanna Wallach}.}
  \bibinfo{year}{2019}\natexlab{}.
\newblock \showarticletitle{Improving fairness in machine learning systems:
  What do industry practitioners need?}. In
  \bibinfo{booktitle}{\emph{Proceedings of the 2019 CHI conference on human
  factors in computing systems}}. \bibinfo{pages}{1--16}.
\newblock


\bibitem[\protect\citeauthoryear{Lakshmanan, Robinson, and Munn}{Lakshmanan
  et~al\mbox{.}}{2020}]%
        {lakshmanan2020machine}
\bibfield{author}{\bibinfo{person}{Valliappa Lakshmanan}, \bibinfo{person}{Sara
  Robinson}, {and} \bibinfo{person}{Michael Munn}.}
  \bibinfo{year}{2020}\natexlab{}.
\newblock \bibinfo{booktitle}{\emph{Machine learning design patterns}}.
\newblock \bibinfo{publisher}{O'Reilly Media}.
\newblock


\bibitem[\protect\citeauthoryear{Lipton}{Lipton}{2018}]%
        {lipton2018mythos}
\bibfield{author}{\bibinfo{person}{Zachary~C Lipton}.}
  \bibinfo{year}{2018}\natexlab{}.
\newblock \showarticletitle{The Mythos of Model Interpretability: In machine
  learning, the concept of interpretability is both important and slippery.}
\newblock \bibinfo{journal}{\emph{Queue}} \bibinfo{volume}{16},
  \bibinfo{number}{3} (\bibinfo{year}{2018}), \bibinfo{pages}{31--57}.
\newblock


\bibitem[\protect\citeauthoryear{Liu, Anguelov, Erhan, Szegedy, Reed, Fu, and
  Berg}{Liu et~al\mbox{.}}{2016}]%
        {liu2016ssdV1}
\bibfield{author}{\bibinfo{person}{Wei Liu}, \bibinfo{person}{Dragomir
  Anguelov}, \bibinfo{person}{Dumitru Erhan}, \bibinfo{person}{Christian
  Szegedy}, \bibinfo{person}{Scott Reed}, \bibinfo{person}{Cheng-Yang Fu},
  {and} \bibinfo{person}{Alexander~C Berg}.} \bibinfo{year}{2016}\natexlab{}.
\newblock \showarticletitle{{SSD}: Single shot multibox detector}. In
  \bibinfo{booktitle}{\emph{European Conference on Computer Vision}}.
  \bibinfo{pages}{21--37}.
\newblock


\bibitem[\protect\citeauthoryear{Ludewig}{Ludewig}{2003}]%
        {ludewig2003models}
\bibfield{author}{\bibinfo{person}{Jochen Ludewig}.}
  \bibinfo{year}{2003}\natexlab{}.
\newblock \showarticletitle{Models in software engineering--an introduction}.
\newblock \bibinfo{journal}{\emph{Software and Systems Modeling}}
  \bibinfo{volume}{2}, \bibinfo{number}{1} (\bibinfo{year}{2003}),
  \bibinfo{pages}{5--14}.
\newblock


\bibitem[\protect\citeauthoryear{Meng and Amalathas}{Meng and
  Amalathas}{2019}]%
        {meng2019new}
\bibfield{author}{\bibinfo{person}{Wong~Hoo Meng} {and}
  \bibinfo{person}{Sagaya~Sabestinal Amalathas}.}
  \bibinfo{year}{2019}\natexlab{}.
\newblock \showarticletitle{A new approach towards developing a prescriptive
  analytical logic model for software application error analysis}. In
  \bibinfo{booktitle}{\emph{Proceedings of the Computational Methods in Systems
  and Software}}. Springer, \bibinfo{pages}{256--274}.
\newblock


\bibitem[\protect\citeauthoryear{Miceli, Yang, Naudts, Schuessler, Serbanescu,
  and Hanna}{Miceli et~al\mbox{.}}{2021}]%
        {miceli2021documenting}
\bibfield{author}{\bibinfo{person}{Milagros Miceli}, \bibinfo{person}{Tianling
  Yang}, \bibinfo{person}{Laurens Naudts}, \bibinfo{person}{Martin Schuessler},
  \bibinfo{person}{Diana Serbanescu}, {and} \bibinfo{person}{Alex Hanna}.}
  \bibinfo{year}{2021}\natexlab{}.
\newblock \showarticletitle{Documenting Computer Vision Datasets: An Invitation
  to Reflexive Data Practices}. In \bibinfo{booktitle}{\emph{Proceedings of the
  2021 ACM Conference on Fairness, Accountability, and Transparency}}.
  \bibinfo{pages}{161--172}.
\newblock


\bibitem[\protect\citeauthoryear{Microsoft}{Microsoft}{2021}]%
        {Microsoft2021HAX}
\bibfield{author}{\bibinfo{person}{Microsoft}.}
  \bibinfo{year}{2021}\natexlab{}.
\newblock \showarticletitle{The HAX Toolkit}.
\newblock  (\bibinfo{year}{2021}).
\newblock
\urldef\tempurl%
\url{https://www.microsoft.com/en-us/haxtoolkit/}
\showURL{%
\tempurl}


\bibitem[\protect\citeauthoryear{Mitchell, Wu, Zaldivar, Barnes, Vasserman,
  Hutchinson, Spitzer, Raji, and Gebru}{Mitchell et~al\mbox{.}}{2019}]%
        {mitchell2019model}
\bibfield{author}{\bibinfo{person}{Margaret Mitchell}, \bibinfo{person}{Simone
  Wu}, \bibinfo{person}{Andrew Zaldivar}, \bibinfo{person}{Parker Barnes},
  \bibinfo{person}{Lucy Vasserman}, \bibinfo{person}{Ben Hutchinson},
  \bibinfo{person}{Elena Spitzer}, \bibinfo{person}{Inioluwa~Deborah Raji},
  {and} \bibinfo{person}{Timnit Gebru}.} \bibinfo{year}{2019}\natexlab{}.
\newblock \showarticletitle{Model cards for model reporting}. In
  \bibinfo{booktitle}{\emph{Proceedings of the conference on fairness,
  accountability, and transparency}}. \bibinfo{pages}{220--229}.
\newblock


\bibitem[\protect\citeauthoryear{Nosek, Spies, and Motyl}{Nosek
  et~al\mbox{.}}{2012}]%
        {nosek2012scientific}
\bibfield{author}{\bibinfo{person}{Brian~A Nosek}, \bibinfo{person}{Jeffrey~R
  Spies}, {and} \bibinfo{person}{Matt Motyl}.} \bibinfo{year}{2012}\natexlab{}.
\newblock \showarticletitle{Scientific utopia: II. Restructuring incentives and
  practices to promote truth over publishability}.
\newblock \bibinfo{journal}{\emph{Perspectives on Psychological Science}}
  \bibinfo{volume}{7}, \bibinfo{number}{6} (\bibinfo{year}{2012}),
  \bibinfo{pages}{615--631}.
\newblock


\bibitem[\protect\citeauthoryear{PAIR}{PAIR}{2019}]%
        {Google2019Guidebook}
\bibfield{author}{\bibinfo{person}{Google PAIR}.}
  \bibinfo{year}{2019}\natexlab{}.
\newblock \showarticletitle{People + AI Guidebook}.
\newblock  (\bibinfo{year}{2019}).
\newblock
\urldef\tempurl%
\url{https://pair.withgoogle.com/guidebook}
\showURL{%
\tempurl}


\bibitem[\protect\citeauthoryear{Parmar, Zhang, and Zhu}{Parmar
  et~al\mbox{.}}{2021}]%
        {parmar2021buggy}
\bibfield{author}{\bibinfo{person}{Gaurav Parmar}, \bibinfo{person}{Richard
  Zhang}, {and} \bibinfo{person}{Jun-Yan Zhu}.}
  \bibinfo{year}{2021}\natexlab{}.
\newblock \showarticletitle{On Buggy Resizing Libraries and Surprising
  Subtleties in FID Calculation}.
\newblock \bibinfo{journal}{\emph{arXiv preprint arXiv:2104.11222}}
  (\bibinfo{year}{2021}).
\newblock


\bibitem[\protect\citeauthoryear{Paszke, Gross, Massa, Lerer, Bradbury, Chanan,
  Killeen, Lin, Gimelshein, et~al\mbox{.}}{Paszke et~al\mbox{.}}{2019}]%
        {Paszke2019PyTorch}
\bibfield{author}{\bibinfo{person}{Adam Paszke}, \bibinfo{person}{Sam Gross},
  \bibinfo{person}{Francisco Massa}, \bibinfo{person}{Adam Lerer},
  \bibinfo{person}{James Bradbury}, \bibinfo{person}{Gregory Chanan},
  \bibinfo{person}{Trevor Killeen}, \bibinfo{person}{Zeming Lin},
  \bibinfo{person}{Natalia Gimelshein}, {et~al\mbox{.}}}
  \bibinfo{year}{2019}\natexlab{}.
\newblock \showarticletitle{{PyTorch}: An Imperative Style, High-Performance
  Deep Learning Library}. In \bibinfo{booktitle}{\emph{Advances in Neural
  Information Processing Systems (NeurIPS)}}. \bibinfo{pages}{8024--8035}.
\newblock


\bibitem[\protect\citeauthoryear{Pineau, Vincent-Lamarre, Sinha, Larivi{\`e}re,
  Beygelzimer, d’Alch{\'e} Buc, Fox, and Larochelle}{Pineau
  et~al\mbox{.}}{2021}]%
        {pineau2021improving}
\bibfield{author}{\bibinfo{person}{Joelle Pineau}, \bibinfo{person}{Philippe
  Vincent-Lamarre}, \bibinfo{person}{Koustuv Sinha}, \bibinfo{person}{Vincent
  Larivi{\`e}re}, \bibinfo{person}{Alina Beygelzimer},
  \bibinfo{person}{Florence d’Alch{\'e} Buc}, \bibinfo{person}{Emily Fox},
  {and} \bibinfo{person}{Hugo Larochelle}.} \bibinfo{year}{2021}\natexlab{}.
\newblock \showarticletitle{Improving reproducibility in machine learning
  research: a report from the NeurIPS 2019 reproducibility program}.
\newblock \bibinfo{journal}{\emph{Journal of Machine Learning Research}}
  \bibinfo{volume}{22} (\bibinfo{year}{2021}).
\newblock


\bibitem[\protect\citeauthoryear{Press, Flannery, Teukolsky, Vetterling,
  et~al\mbox{.}}{Press et~al\mbox{.}}{1989}]%
        {press1989numerical}
\bibfield{author}{\bibinfo{person}{William~H Press}, \bibinfo{person}{Brian~P
  Flannery}, \bibinfo{person}{Saul~A Teukolsky}, \bibinfo{person}{William~T
  Vetterling}, {et~al\mbox{.}}} \bibinfo{year}{1989}\natexlab{}.
\newblock \bibinfo{title}{Numerical recipes}.
\newblock
\newblock


\bibitem[\protect\citeauthoryear{Radford, Wu, Child, Luan, Amodei, and
  Sutskever}{Radford et~al\mbox{.}}{2019}]%
        {radford2019language}
\bibfield{author}{\bibinfo{person}{Alec Radford}, \bibinfo{person}{Jeff Wu},
  \bibinfo{person}{Rewon Child}, \bibinfo{person}{David Luan},
  \bibinfo{person}{Dario Amodei}, {and} \bibinfo{person}{Ilya Sutskever}.}
  \bibinfo{year}{2019}\natexlab{}.
\newblock \showarticletitle{Language Models are Unsupervised Multitask
  Learners}.
\newblock  (\bibinfo{year}{2019}).
\newblock


\bibitem[\protect\citeauthoryear{Raff}{Raff}{2019}]%
        {raff2019step}
\bibfield{author}{\bibinfo{person}{Edward Raff}.}
  \bibinfo{year}{2019}\natexlab{}.
\newblock \showarticletitle{A step toward quantifying independently
  reproducible machine learning research}.
\newblock \bibinfo{journal}{\emph{Advances in Neural Information Processing
  Systems}}  \bibinfo{volume}{32} (\bibinfo{year}{2019}),
  \bibinfo{pages}{5485--5495}.
\newblock


\bibitem[\protect\citeauthoryear{Raghavan and Chand}{Raghavan and
  Chand}{1989}]%
        {raghavan1989diffusing}
\bibfield{author}{\bibinfo{person}{Sridhar~A. Raghavan} {and}
  \bibinfo{person}{Donald~R. Chand}.} \bibinfo{year}{1989}\natexlab{}.
\newblock \showarticletitle{Diffusing software-engineering methods}.
\newblock \bibinfo{journal}{\emph{IEEE software}} \bibinfo{volume}{6},
  \bibinfo{number}{4} (\bibinfo{year}{1989}), \bibinfo{pages}{81--90}.
\newblock


\bibitem[\protect\citeauthoryear{Raji and Yang}{Raji and Yang}{2019}]%
        {raji2019ml}
\bibfield{author}{\bibinfo{person}{I.~D. Raji} {and} \bibinfo{person}{J.
  Yang}.} \bibinfo{year}{2019}\natexlab{}.
\newblock \showarticletitle{ABOUT ML: Annotation and benchmarking on
  understanding and transparency of machine learning lifecycles}.
\newblock \bibinfo{journal}{\emph{arXiv preprint arXiv:1912.06166}}
  (\bibinfo{year}{2019}).
\newblock


\bibitem[\protect\citeauthoryear{Sculley, Holt, Golovin, Davydov, Phillips,
  Ebner, Chaudhary, Young, Crespo, and Dennison}{Sculley et~al\mbox{.}}{2015}]%
        {sculley2015hidden}
\bibfield{author}{\bibinfo{person}{David Sculley}, \bibinfo{person}{Gary Holt},
  \bibinfo{person}{Daniel Golovin}, \bibinfo{person}{Eugene Davydov},
  \bibinfo{person}{Todd Phillips}, \bibinfo{person}{Dietmar Ebner},
  \bibinfo{person}{Vinay Chaudhary}, \bibinfo{person}{Michael Young},
  \bibinfo{person}{Jean-Francois Crespo}, {and} \bibinfo{person}{Dan
  Dennison}.} \bibinfo{year}{2015}\natexlab{}.
\newblock \showarticletitle{Hidden technical debt in machine learning systems}.
\newblock \bibinfo{journal}{\emph{Advances in neural information processing
  systems}}  \bibinfo{volume}{28} (\bibinfo{year}{2015}),
  \bibinfo{pages}{2503--2511}.
\newblock


\bibitem[\protect\citeauthoryear{Shibui}{Shibui}{2020}]%
        {Shibui2020patterns}
\bibfield{author}{\bibinfo{person}{Y. Shibui}.}
  \bibinfo{year}{2020}\natexlab{}.
\newblock \showarticletitle{Machine learning system design patterns}.
\newblock  (\bibinfo{year}{2020}).
\newblock
\urldef\tempurl%
\url{https://github.com/mercari/ml-system-design-pattern}
\showURL{%
\tempurl}


\bibitem[\protect\citeauthoryear{Steegen, Tuerlinckx, Gelman, and
  Vanpaemel}{Steegen et~al\mbox{.}}{2016}]%
        {steegen2016increasing}
\bibfield{author}{\bibinfo{person}{Sara Steegen}, \bibinfo{person}{Francis
  Tuerlinckx}, \bibinfo{person}{Andrew Gelman}, {and} \bibinfo{person}{Wolf
  Vanpaemel}.} \bibinfo{year}{2016}\natexlab{}.
\newblock \showarticletitle{Increasing transparency through a multiverse
  analysis}.
\newblock \bibinfo{journal}{\emph{Perspectives on Psychological Science}}
  \bibinfo{volume}{11}, \bibinfo{number}{5} (\bibinfo{year}{2016}),
  \bibinfo{pages}{702--712}.
\newblock


\bibitem[\protect\citeauthoryear{Washizaki, Khomh, Gu{\'e}h{\'e}neuc, Takeuchi,
  Natori, Doi, and Okuda}{Washizaki et~al\mbox{.}}{2022}]%
        {washizaki2022software}
\bibfield{author}{\bibinfo{person}{Hironori Washizaki}, \bibinfo{person}{Foutse
  Khomh}, \bibinfo{person}{Yann-Ga{\"e}l Gu{\'e}h{\'e}neuc},
  \bibinfo{person}{Hironori Takeuchi}, \bibinfo{person}{Naotake Natori},
  \bibinfo{person}{Takuo Doi}, {and} \bibinfo{person}{Satoshi Okuda}.}
  \bibinfo{year}{2022}\natexlab{}.
\newblock \showarticletitle{Software Engineering Design Patterns for Machine
  Learning Applications}.
\newblock \bibinfo{journal}{\emph{IEEE Computer}} \bibinfo{volume}{55},
  \bibinfo{number}{3} (\bibinfo{year}{2022}), \bibinfo{pages}{1--9}.
\newblock


\bibitem[\protect\citeauthoryear{Washizaki, Uchida, Khomh, and
  Gu{\'e}h{\'e}neuc}{Washizaki et~al\mbox{.}}{2019}]%
        {washizaki2019studying}
\bibfield{author}{\bibinfo{person}{Hironori Washizaki}, \bibinfo{person}{Hiromu
  Uchida}, \bibinfo{person}{Foutse Khomh}, {and} \bibinfo{person}{Yann-Ga{\"e}l
  Gu{\'e}h{\'e}neuc}.} \bibinfo{year}{2019}\natexlab{}.
\newblock \showarticletitle{Studying software engineering patterns for
  designing machine learning systems}. In \bibinfo{booktitle}{\emph{2019 10th
  International Workshop on Empirical Software Engineering in Practice
  (IWESEP)}}. IEEE, \bibinfo{pages}{49--495}.
\newblock


\bibitem[\protect\citeauthoryear{Wattenberg, Vi{\'e}gas, and
  Johnson}{Wattenberg et~al\mbox{.}}{2016}]%
        {wattenberg2016use}
\bibfield{author}{\bibinfo{person}{Martin Wattenberg},
  \bibinfo{person}{Fernanda Vi{\'e}gas}, {and} \bibinfo{person}{Ian Johnson}.}
  \bibinfo{year}{2016}\natexlab{}.
\newblock \showarticletitle{How to use t-SNE effectively}.
\newblock \bibinfo{journal}{\emph{Distill}} \bibinfo{volume}{1},
  \bibinfo{number}{10} (\bibinfo{year}{2016}), \bibinfo{pages}{e2}.
\newblock


\bibitem[\protect\citeauthoryear{Weller}{Weller}{2019}]%
        {weller2019transparency}
\bibfield{author}{\bibinfo{person}{Adrian Weller}.}
  \bibinfo{year}{2019}\natexlab{}.
\newblock \showarticletitle{Transparency: motivations and challenges}.
\newblock In \bibinfo{booktitle}{\emph{Explainable AI: Interpreting, Explaining
  and Visualizing Deep Learning}}. \bibinfo{publisher}{Springer},
  \bibinfo{pages}{23--40}.
\newblock


\bibitem[\protect\citeauthoryear{Wong, Amalathas, and Zitkova}{Wong
  et~al\mbox{.}}{2019}]%
        {wong2019prescriptive}
\bibfield{author}{\bibinfo{person}{Hoo~Meng Wong},
  \bibinfo{person}{Sagaya~Sabestinal Amalathas}, {and} \bibinfo{person}{Tatana
  Zitkova}.} \bibinfo{year}{2019}\natexlab{}.
\newblock \showarticletitle{A prescriptive logic model for software application
  root cause analysis}.
\newblock \bibinfo{journal}{\emph{European Journal of Electrical Engineering
  and Computer Science}} \bibinfo{volume}{3}, \bibinfo{number}{5}
  (\bibinfo{year}{2019}).
\newblock


\bibitem[\protect\citeauthoryear{Yokoyama}{Yokoyama}{2019}]%
        {yokoyama2019machine}
\bibfield{author}{\bibinfo{person}{Haruki Yokoyama}.}
  \bibinfo{year}{2019}\natexlab{}.
\newblock \showarticletitle{Machine learning system architectural pattern for
  improving operational stability}. In \bibinfo{booktitle}{\emph{2019 IEEE
  International Conference on Software Architecture Companion (ICSA-C)}}. IEEE,
  \bibinfo{pages}{267--274}.
\newblock


\bibitem[\protect\citeauthoryear{Yuan, Alsallakh, Kokhlikyan, Miglani, and
  Reblitz-Richardson}{Yuan et~al\mbox{.}}{2021}]%
        {yuan2021convolution}
\bibfield{author}{\bibinfo{person}{J. Yuan}, \bibinfo{person}{B. Alsallakh},
  \bibinfo{person}{N. Kokhlikyan}, \bibinfo{person}{V. Miglani}, {and}
  \bibinfo{person}{O. Reblitz-Richardson}.} \bibinfo{year}{2021}\natexlab{}.
\newblock \showarticletitle{Convolution Can Incur Foveation Effects}. In
  \bibinfo{booktitle}{\emph{Beyond static papers: Rethinking how we share
  scientific understanding in ML-ICLR 2021 workshop}}.
\newblock


\bibitem[\protect\citeauthoryear{Zoph, Ghiasi, Lin, Cui, Liu, Cubuk, and
  Le}{Zoph et~al\mbox{.}}{2020}]%
        {zoph2020rethinking}
\bibfield{author}{\bibinfo{person}{Barret Zoph}, \bibinfo{person}{Golnaz
  Ghiasi}, \bibinfo{person}{Tsung-Yi Lin}, \bibinfo{person}{Yin Cui},
  \bibinfo{person}{Hanxiao Liu}, \bibinfo{person}{Ekin~D Cubuk}, {and}
  \bibinfo{person}{Quoc~V Le}.} \bibinfo{year}{2020}\natexlab{}.
\newblock \showarticletitle{Rethinking pre-training and self-training}.
\newblock \bibinfo{journal}{\emph{arXiv preprint arXiv:2006.06882}}
  (\bibinfo{year}{2020}).
\newblock


\end{thebibliography}

\appendix


\clearpage

\begin{figure*}[th!]

\begin{center}
   Method Card - Traffic Light Detection
\end{center} 
\hspace{1mm}
\begin{mdframed}
\begin{multicols}{2}

\textbf{ Basic Method Information}
\hrulefill
\paragraph{\textbf{Name, version, and application domain(s):}}
Traffic light detection based on SSD, V1, for street scene analysis.
\paragraph{\textbf{Method purpose and appropriate uses:}}
Detect rectangular traffic-light instances in a street scene, represented by a monocular camera image.
The detector can be used in offline analysis, e.g., to support street scene retrieval based on traffic criteria and in  similar traffic analysis tasks.
\paragraph{\textbf{Method definition:}} 
Traffic light detection based on the SSD object detection architecture.
Published in~\cite{behrendt2017deep}, reference implementation in \url{https://github.com/bosch-ros-pkg/bstld}.
\paragraph{\textbf{Example input and output:}}
The Output is a list of bounding boxes representing the detected objects, along with detection score and detected traffic light color.
\newline
\newline
\phantom{ph} \includegraphics[width=0.9\linewidth]{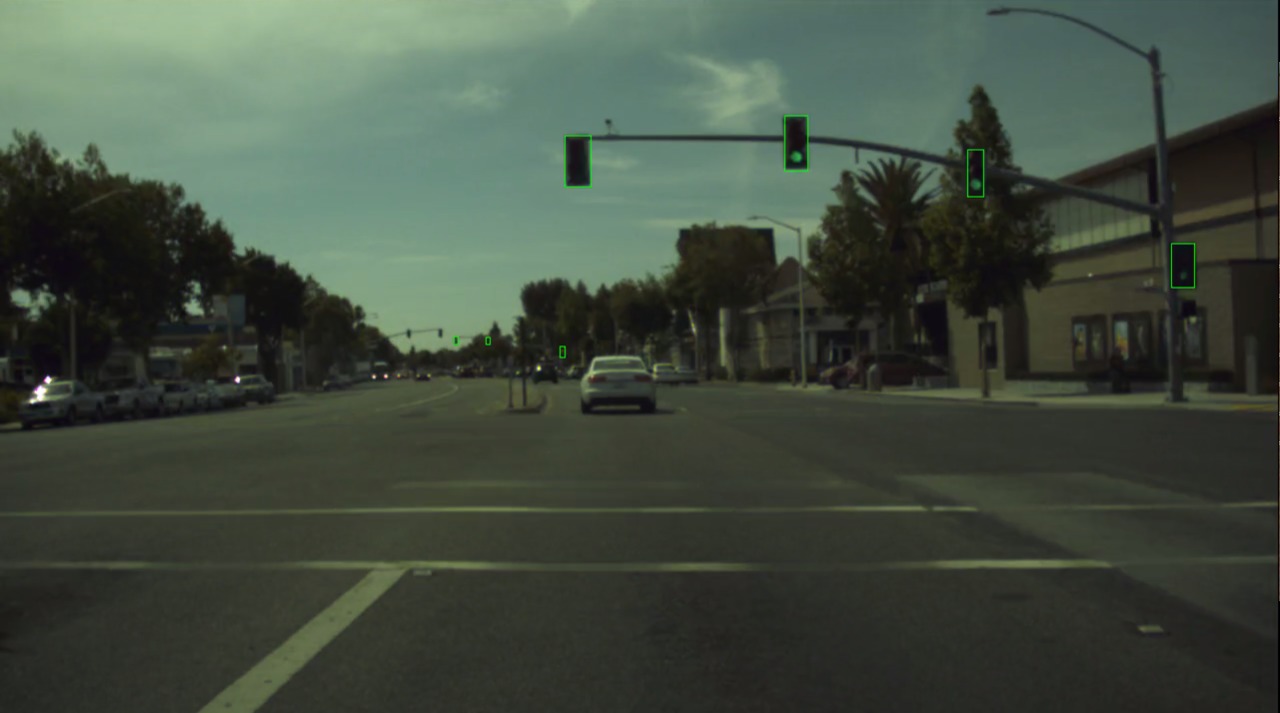} \\

\textbf{Safety and Troubleshooting}
\hrulefill
\paragraph{\textbf{Inappropriate use and common usage pitfalls:}}
\begin{itemize}
    \item Must NOT be used in autonomous driving, it is not designed to handle all traffic scenarios.
    \item Should not be used with night-time scenes. 
\end{itemize}

\paragraph{\textbf{Known weaknesses and biases:}} 
\begin{itemize}
\item Can only handle rectangular bounding boxes during training and inference.
\item Biased against red traffic lights as they are very similar to vehicle tail lights.
\end{itemize}

\paragraph{\textbf{How to detect biases in the model internals:}}
\begin{itemize}
\item Feed a zero input and look for activation artifacts
\item Average the feature maps over random inputs.
\item Is the average kernel per conv. layer symmetric?
\end{itemize}
\paragraph{\textbf{Common failure modes, root causes, and mitigation:}}
\begin{itemize}
\item Red traffic lights likely undetected if in proximity of vehicles (will likely be deemed as tail lights). Consider generating training samples with such scenarios, or first applying a car detector and masking detected cars.
\item Green traffic lights with vegetation background.
Consider using a backbone that has higher shape bias and lower texture bias~\cite{geirhos2018imagenettrained}.
\item Traffic lights with street ads in the background.
Consider augmentation with generating data.
\end{itemize}

\textbf{Data Preparation}
\hrulefill
\paragraph{\textbf{Input and output format, shape, and data type:}}
\begin{itemize}
\item The input is a $1280 \times 840$ RGB image, with integer intensity values in the range $[0, 225]$. Do not resize the image as this can compromise the aspect ratio of the traffic lights.
\item The output is a ranked list of detected traffic lights. 
Each item in the list contains the bounding box $((x1, y1), (x2, y2))$ in a relative coordinate system $[0, 1] \times [0, 1]$, and the production confidence normalized to $[0, 1]$.
\end{itemize}

\paragraph{\textbf{Data transformation and normalization:}}
The input image is first normalized to the range $[0, 1]$.

\paragraph{\textbf{Recommended sampling and balancing:}}
\begin{itemize}
\item The frames in available datasets are captured at a constant interval. Hence, scenes at red traffic lights tend to be almost identical and over-represented. Consider similarity-based sampling. 
\item Yellow traffic lights are limited in available datasets and often have a larger size as they tend to be captured close to intersections. Data augmentation should focus on increasing their frequency.
\end{itemize}

\paragraph{\textbf{Recommended batching scheme and batch size.}}
\begin{itemize}
\item Use a batch size of 4 or higher if the GPU allows.
\item Ensure that batches contain diverse scenes instead of consecutive frames.
\item Use  \texttt{InstanceNormalization} for a batch size of 1 or 2 and \texttt{BatchNorm} with batch size of 4.
\end{itemize}

\paragraph{\textbf{Required data augmentation and shuffling.}}
\begin{itemize}
\item Make sure to enable horizontal image flipping.
\item Shuffle individual frames instead of batches.
\end{itemize}

\paragraph{\textbf{Validation and train-test splitting schemes:}}
Ensure sufficient diversity of the different splits, especially given the sequential nature of the collected scenes. Ensure that yellow traffic lights are sufficiently present in the validation set as they are likely a minority.

\end{multicols}
\hspace{1mm}
\end{mdframed}

 \label{fig:ex1_method_card} 
\end{figure*}

\clearpage

\begin{figure*}[th!]

\begin{center}
   Method Card - Traffic Light Detection (continued)
\end{center} 
\hspace{1mm}

\begin{mdframed}
\begin{multicols}{2}

\textbf{Modelling and Optimization}
\hrulefill
\paragraph{\textbf{Architecture family and components:}}
\begin{itemize}
\item The SSD multi-box architecture~\cite{liu2016ssdV1} with a box predictor for each traffic light color.
\item The backbone CNN can be a ResNet, a MobileNet, EfficientNet or another architecture depending on available computing capacity.
\item Avoid backbones that use dilation.
\item Use maxpooling instead of strided convolution.
\end{itemize}

\paragraph{\textbf{Hyperparameters, applicable values, known impact:}}
\begin{itemize}
\item Use mirror padding to reduce artifacts. 0-padding can create blind spots with small objects.
\item Resize each input dimension to a $a \cdot 2^d + 1$ where $d$ is the number of pooling layers an $a$ is a scalar. This avoids asymmetric filters.
\end{itemize}

\paragraph{\textbf{Training objective(s), and optimizer(s):}}
Use
\begin{itemize}
\item \texttt{weighted sigmoid} as classification loss and \texttt{weighted smooth L1} for localization loss.
\item Hard negative miner and NMS to reduce FPs.
\item {RMSProp} optimizer with MobileNet and \texttt{SGD with Momentum} for ResNet backbone.
\end{itemize}

\paragraph{\textbf{Parameter initialization:}} Use Xavier's method if training from scratch. If finetuning, use a checkpoint trained with self-supervision.

\paragraph{\textbf{Regularization scheme, capacity selection:}}
We do NOT recommend using dropout in the box-predictor of the SSD.
We recommend an L2 regularizer with a factor of 0.00004.

\paragraph{\textbf{Weight quantization, recommended bit depth:}}
Given the small size of the target objects, we recommend 32-bit FP values to encode the weights without quantization.

\paragraph{\textbf{Parallelization at training and inference time:}}
We recommend identical GPUs to parallelize the training, and ensuring the batches are distributed randomly across them.

\paragraph{\textbf{Recommended model compression techniques:}}
Given the small size of the target objects, compression can potentially wash out fundamental features.
\newline
\newline
\newline
\newline
\newline

\textbf{Method Benchmarking}
\hrulefill
\paragraph{\textbf{Performance metric(s) and applicable threshold(s):}}
\begin{itemize}
\item Use Average Precision (AP) to compare models  overall, per light category, and per object size.
\item Use the IoU overlap threshold to determine if a detected object covers the ground truth instance.
\item A score threshold defines the operating point.
\end{itemize}
\paragraph{\textbf{Threshold selection:}}
Calibrate the IoU and score thresholds based on the intended use.
If using lower thresholds for higher recall, consider human reviewers to double check.

\paragraph{\textbf{Overfitting detection:}} Standard analysis on training and test set performance.

\paragraph{\textbf{Training and inference time efficiency:}}
SSDs are generally fast both at training and at inference.
A MobileNet backbone converges within hours on BSTLD, and can perform detection in $38ms$ on a single-core machine.

\paragraph{\textbf{Available benchmarks:}}
Results on BSTLD: \url{https://github.com/bosch-ros-pkg/bstld\#results}
\\

\textbf{Interpretability and Explainability}
\hrulefill
\paragraph{\textbf{Applicable feature attribution methods:}}
\texttt{GradCAM}, \texttt{LIME} (applied to super-pixels), and window occlusion help determine scene features involved in the detection.

\paragraph{\textbf{How to identify influential training instances:}}
Use algorithms based on comparing embeddings, such as \texttt{TracIn}, to identify proponents and opponents.  

\paragraph{\textbf{How to identify internal concepts and features learned:}}
Use \texttt{TCAV} when example scenes for street concepts are available.
Use \texttt{NetDissect} when scenes with segmentation masks are available.
\\

\textbf{Robustness}
\hrulefill
\paragraph{\textbf{Out-of-distribution behavior:}}
The method is likely to depend on regional street features, and hence to underperform on scenes from cities outside of the training set.
Also, distorted or significantly different tones of green, yellow, or red will likely be undetected.
\paragraph{\textbf{Detecting and mitigating data and model drifts:}}
Traffic lights are constantly being updated, and hence the training data need to be regularly updated. 
Novel objects might resemble traffic lights, incurring new false positives.

\end{multicols}
\hspace{1mm}
\end{mdframed}

 \label{fig:ex1_method_card_ctnd} 
\end{figure*}

\end{document}